\documentclass[journal=nanoletters,manuscript=article]{achemso}

\usepackage{chemformula} % Formula subscripts using \ch{}
\usepackage[T1]{fontenc} % Use modern font encodings

\author{Heeyuen Koh}
\email{heeyuen.koh@gmail.com}
\affiliation[Soft Foundry Institute]
{Soft Foundry Institute, Seoul National University, 1 Gwanak-ro, Gwanak-gu, Seoul, 08826, Korea}

\author{Jae Young Lee}
\affiliation
{Department of Mechanical Engineering, Ajou University, 206 World cup-ro, Yengtong-gu, Suwon, Gyeonggi-do, 16499, Korea}
\author{Jae Gyung Lee}
\affiliation
{Department of Mechanical and Aerospace Engineering,
Seoul National University, 1 Gwanak-ro, Gwanak-gu, Seoul, 08826, Korea}
\title  {Bend-Twist Coupling and its Design from Geometrical Constraints in Double-Stranded DNA}

\keywords{Nonlinearity, Supercoiling, Nanoscale helical structure, DNA, Bend-twist coupling, Coarse-grained simulation}

\begin{document}

%% require that it be printed on the abstract page. It will
\begin{tocentry}

\includegraphics[scale=0.45, trim={0 300 300 0},clip]{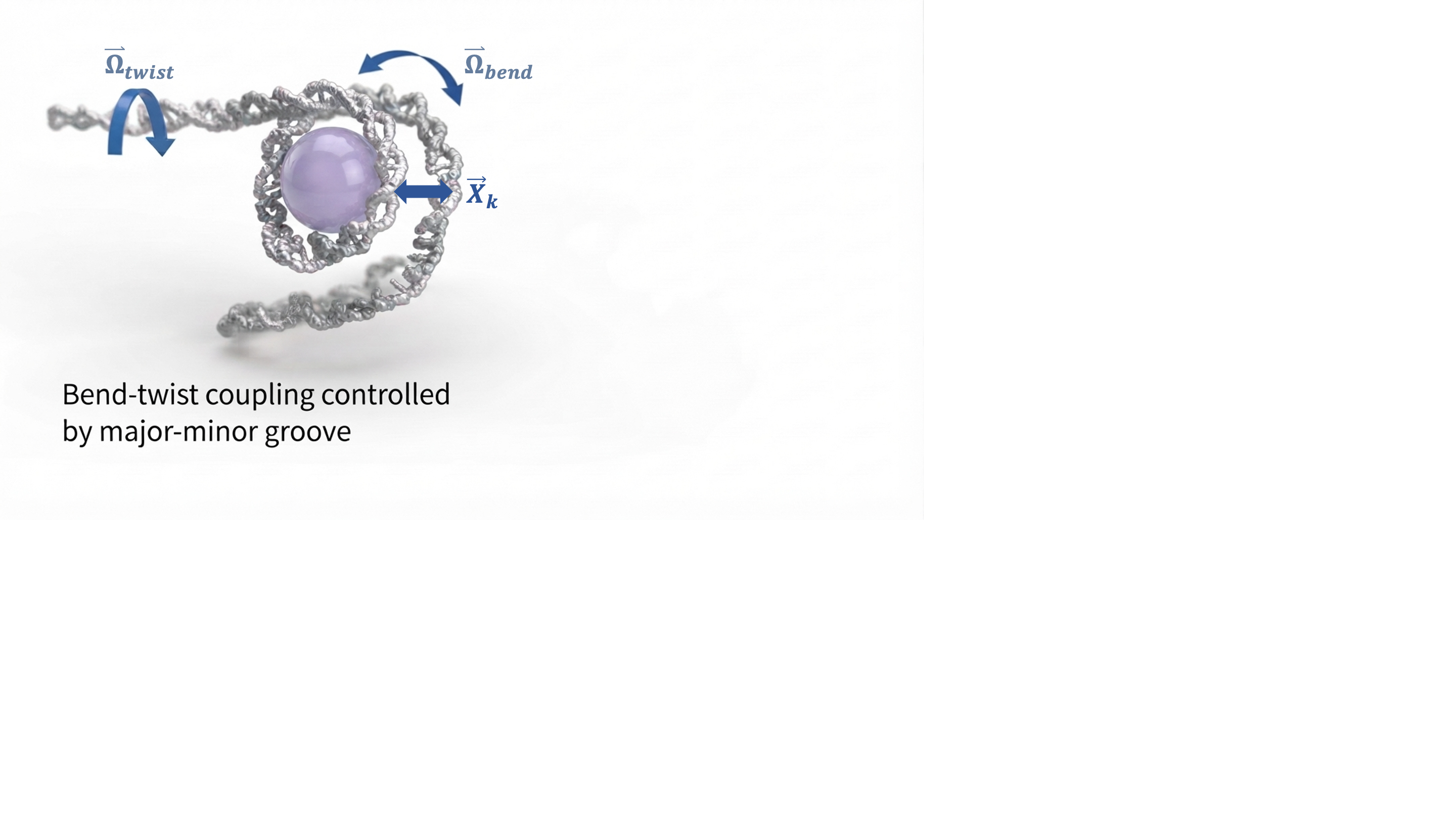}

\end{tocentry}

\begin{abstract}
The versatility of helical structures in nature indicates their propensity to yield useful functions that conserve, propagate, and assemble. However, it remains unclear precisely why these geometric forms benefit their functions. The structural features of the B-DNA strand, particularly the major and minor grooves and their relation to bend-twist coupling and associated dynamics, serve as a compelling example. Uncovering the direct connection between molecular geometry and induced dynamics is fundamental to bridging related functionality, while unlocking design principles for various helical nanoscale applications. In this paper, we leverage the geometric features of helicity to derive an ansatz for the bend-twist coupling of a curved strand. The constraints of the ansatz reveal how major and minor grooves shape a curved strand into a quintessential structural motif, imparting robust stability essential for DNA packaging. Coarse-grained molecular dynamics simulations validate the analytical results and highlight sequence-dependent preferences during the wrapping process.  
\end{abstract}

%\date{December 2018}%
\maketitle

%\tableofcontents

%\section{Introduction}
The various functions of helical strands in numerous studies reflect a wide range of dynamics and highlight how efficiently and effectively helicity-driven dynamics operate in nature. For instance, DNA strands encode genetic information via Watson-Crick base pairing. Their unique architecture allows for high-density data packaging, facilitating genetic transmission by enabling the system to reversibly compress and unpack the strand to regulate organelle and cellular function. Likewise, compaction and ejection/extrusion of helical strands in microorganisms\cite{HIRSH2015251, smith2001bacteriophage,coshic2024structure}, climbing tendrils of plants\cite{wang2013hierarchical}, and actin filaments in muscle\cite{enrique2010origin,borovikov2025twisting} rely on the nonlinear dynamics induced by helicity. Such versatility in dynamics is one of the reasons for the recent shift toward embracing helical design in artificial objects, a trend widely pursued for nano/microscale helices\cite{ye2025advanced,yan2025interface} using metals\cite{xue2020mechanically}, semiconductors\cite{cho2023bioinspired}, oxides\cite{dong2022self}, or peptide-based motifs\cite{wang2025chiral,caimi2023twisted}. Establishing design rules that map helical geometry to specific functional dynamics is essential to fully harness the nonlinear characteristics of both natural and synthetic helical structures.

Bend-twist coupled deformation is a fundamental characteristic of helical strands, extensively studied across diverse dynamic contexts. For mechanisms that defy standard linear elasticity and persistence length such as genome compaction in virus capsids\cite{coshic2024structure,farrell2024spool}, nucleosomal DNA assembly\cite{Freeman2014a,Nash2015,Lequieu2016,Zhao2019,Bae2021a,Bae2021b}, and the explosive firing of predatory cnidarian nematocysts, the stinging organelles\cite{karabulut2022architecture}, bend-twist coupling is regarded as the governing factor that drives the underlying dynamics. In these highly confined environments, the coupling dictates helical buckling behaviors\cite{wang2025buckling,farrell2024spool} and twist-induced morphological transitions\cite{HIRSH2015251,karabulut2022architecture}. For artificially designed cases, bend-twist coupling expands the range of possible dynamics in the self-propulsion of linear objects\cite{makanga2026instability,butler2026elastohydrodynamics}, the morphology of DNA origami systems\cite{dietz2009folding,park2025snupi}, artificially designed peptide systems\cite{wang2025chiral}, or robotic systems\cite{xu2025programmable,tani2024soft}. In most studies, the Frenet-Serret formula in the Darboux frame is used to describe energetic characteristics. Yet, the origin of the bend-twist coupling from the Frenet formula\cite{DF1, DF2} pinpoints that bend-twist coupling is fundamentally geometric, not purely the consequence of the energetic landscape from coupling elasticity. Elucidating the structural and geometric etiology of bend-twist coupling guides the transition toward engineering next-generation artificial structures that exploit these nonlinear mechanics.

The bend-twist coupling from the Frenet formula\cite{DF1, DF2} in differential geometry, which is defined in a slightly different context from the Frenet-Serret formula, is apparent when considering a bar with any cross-sectional shape exhibiting constant curvature, which induces differences in the workload between the outer and inner rims of the curved object. It is the twist deformation that equally distributes the deformation rate on each vertex or any point on the circumference of the cross section of a system that is either continuous or connected by units of the discretized structure. In this paper, we introduce an ansatz for the bend-twist coupling induced by double-helical geometry, formulated in accordance with the Frenet formula. The ansatz describes the geometric constraints derived from bend-twist correlation and verifies how the major-minor groove structure of the double-helical DNA strand can lead to the functionality of the DNA dynamics, offering a design rule for artificial helical systems. %  
%The bend-twist coupling ansatz derived from the geometric conditions of the DNA strand is directly applicable to synthetic helical systems as a foundational design principle. 

The stability of tightly curved strands in viral capsids and chromatin systems is a compelling case to demonstrate this cross-system applicability, proving that the ansatz can successfully map intrinsic geometric constraints to functional macro-structural stability. The robustness of DNA supercoiling against environmental fluctuations and in processes such as strand sliding within the wrapped conformation during translocation\cite{armeev2021histone,fedulova2024molecular,sabantsev2019direct}, suggests that the supercoiled DNA\cite{Richmond2003,TJRichmond1984,Garai2015} exhibits an inherent dynamical stability that cannot easily be inferred from protein binding affinity or sequence-dependent elasticity. If geometric features of the helicity drive such stability, it is the constraints imposed by the geometrically defined ansatz that govern the strand's dynamics, inducing a substantial twist deformation of approximately 40 degrees while maintaining bending close to that of nucleosomal DNA \cite{Richmond2003,TJRichmond1984,Garai2015}, and ensuring structural resilience within the 5.5 nm height of the superhelix during strand translocation\cite{armeev2021histone,fedulova2024molecular,sabantsev2019direct}. Deformations against the topological or geometrical stability conferred by the height of an inherent superhelix structure provide further examples of where a geometrical perspective can refine design strategies and address unresolved dynamical stability. Examples include viral capsid strand compaction \cite{coshic2024structure,farrell2024spool}, highly bent DNA strands \cite{Yoo2021,Nash2015,Lequieu2016,Zhao2019,Bae2021a,Bae2021b}, and minicircles \cite{Bernard2003,Zoli2014,Skoruppa2018a,Caraglio2019,Nomidis2019a,Kim2021,Yoo2021,Kim2021} or ring-shaped DNA origami \cite{kim2021programming}, where the inherent bend-twist coupling is constrained by geometric or energetic requirements. Furthermore, the proposed geometrical ansatz also accounts for the accumulation of bending that drives strand folding, as well as the resulting balance of dihedral angles alongside bending angles. These geometrically defined dynamics and their accumulants can serve as design principles and governing factors for both artificial and natural helical systems.

 %Theoretical modeling of the geometrical constraints arising from the given helicity of B-DNA is presented in the Theoretical Basis and Models section. The stability of 1.7 turns of the superhelix, determined from the geometrical constraint at base-pair resolution, is explained in the Results. A more detailed explanation of how the ansatz maps stability attributes directly onto the major-minor groove structure is provided in the Discussion section, along with an analysis of the induced energetics. Finally, the paper concludes with a summary of the main findings.

%\section{Theoretical Basis and Models}
\subsection{Geometrical Constraints in Deformable Linear Object }

In engineering models, the Frenet formula provides a geometric representation of a deformable linear object, such as a cord or tube\cite{DF1,DF2,DF3}. A bend-twist coupling of such systems is derived as a function of the curvature of the strand, which is

\begin{eqnarray}\label{eq:eq0}
\omega_{Fr}^2 = \left( \frac{d\phi_e}{ds}\cos\theta_e + \frac{d\psi_e}{ds} \right) ^2.
\end{eqnarray}
 
  $\omega_{Fr}$ is the torsional rate from the Frenet formula. $\phi_e$, $\theta_e$, and $\psi_e$ are Euler angles defined in the Darboux frame, which is a coordinate system defined at a certain point at $s$, which is the arclength of the strand. Because torsional deformation, $\omega_{Fr}$, is defined from the set of angles $\phi_e$ and $\theta_e$, which are from the bending of the strand that forms the curvature, Eq.(\ref{eq:eq0}) becomes a functional relation between bending and torsion. %This geometric constraint among the Euler angles means that the torsional angle arising from strand bending in Eq.(\ref{eq:eq0}) can be regarded as a prerequisite to be considered separately from the coupled elastic moduli.

The Frenet-Serret formula\cite{DF3}, which offers the deformation rate in an orthonormal vector set from the curvature and tangential information of the curved strand from Darboux frame, has been applied to the DNA strand dynamics. Following the previous work\cite{Marko1994}, the deformation rate on each unit vector of the curved strand with inherited helicity is defined as

\begin{eqnarray}\label{eq:eq1}
\frac{d\hat{\textbf e}_i}{ds}  = \left( \vec{\Omega} +\omega_0 \hat{\textbf e}_3 \right) \times \hat{\textbf e}_i.
\end{eqnarray}

$\hat{\textbf e}_{i}, i=1,..,3$ are the unit vectors of the coordinate system, $\{\hat{\textbf e}_i\}_{n}$ on the cross section defined for the n th base pair in the strand. $\hat{\textbf e}_1$ is aligned to make a twofold symmetry of the major-minor groove, and $\hat{\textbf e}_3$ is along the normal vector at the center of the cross section. $\vec{\Omega}$ represents the rotation vector defined at $s$, which is the arclength of the strand. $\hat{\textbf e}_3$ is aligned for the tangent of $s$ as shown in Fig. 1. While $\Omega_1$ and $\Omega_2$ correspond to bending, $\Omega_3$ describes torsion in
$\vec{\Omega} = \Omega_1\hat{\textbf e}_1 +  \Omega_2\hat{\textbf e}_2 +  \Omega_3\hat{\textbf e}_3$. $\omega_0$ is the helicity of the strand, which can be altered by additional twist deformation, $\Omega_3$. Each component of $\vec{\Omega}$ is representing the rotational variable such as tilt ($\tau$), roll ($\rho$) and twist ($\omega$), respectively. To offer more simple and intuitive perspective, we suggested a derivation from geometrical feature of the strand in next subsection. %rather than converting Eq.(\ref{eq:eq1}) into Euler angle notation as a direct implication of Eq.(\ref{eq:eq0}).

\subsection{Geometrically given bend-twist coupling}

Deriving bend-twist coupling from the geometry of the nucleotide is another possible approach under the condition that the cross section of the base pair is rigid enough. According to Eq.(\ref{eq:eq1}), the deformation of any vector defined in coordinate system $\{\hat{\textbf e}_i\}_{n}$ is

\begin{eqnarray}\label{eq:eq2}
d\vec{x} = \sum_i x_i \frac{d\hat{e}_i}{ds} ds = \pmb{\Theta} \vec{x},
\end{eqnarray}
with
\begin{eqnarray}\label{eq:eq2_}
\pmb{\Theta} = \begin{bmatrix} 0 & -(\Omega_3+\omega_0) & \Omega_2 \\ (\Omega_3+\omega_0) & 0 & -\Omega_1  \\ -\Omega_2 & \Omega_1 & 0 \end{bmatrix}ds. % \begin{bmatrix} x_1 \\ x_2 \\ x_3 \end{bmatrix} 
\end{eqnarray}

$\vec{x}$ is an arbitrary vector defined in $\{\hat{\textbf e}_i\}_{n}$, and $\frac{d\vec{x}}{ds}$ is the deformation rate of the vector, $\vec{x}$.

\begin{figure}
\includegraphics[scale=0.45, trim={0 250 350 10},clip]{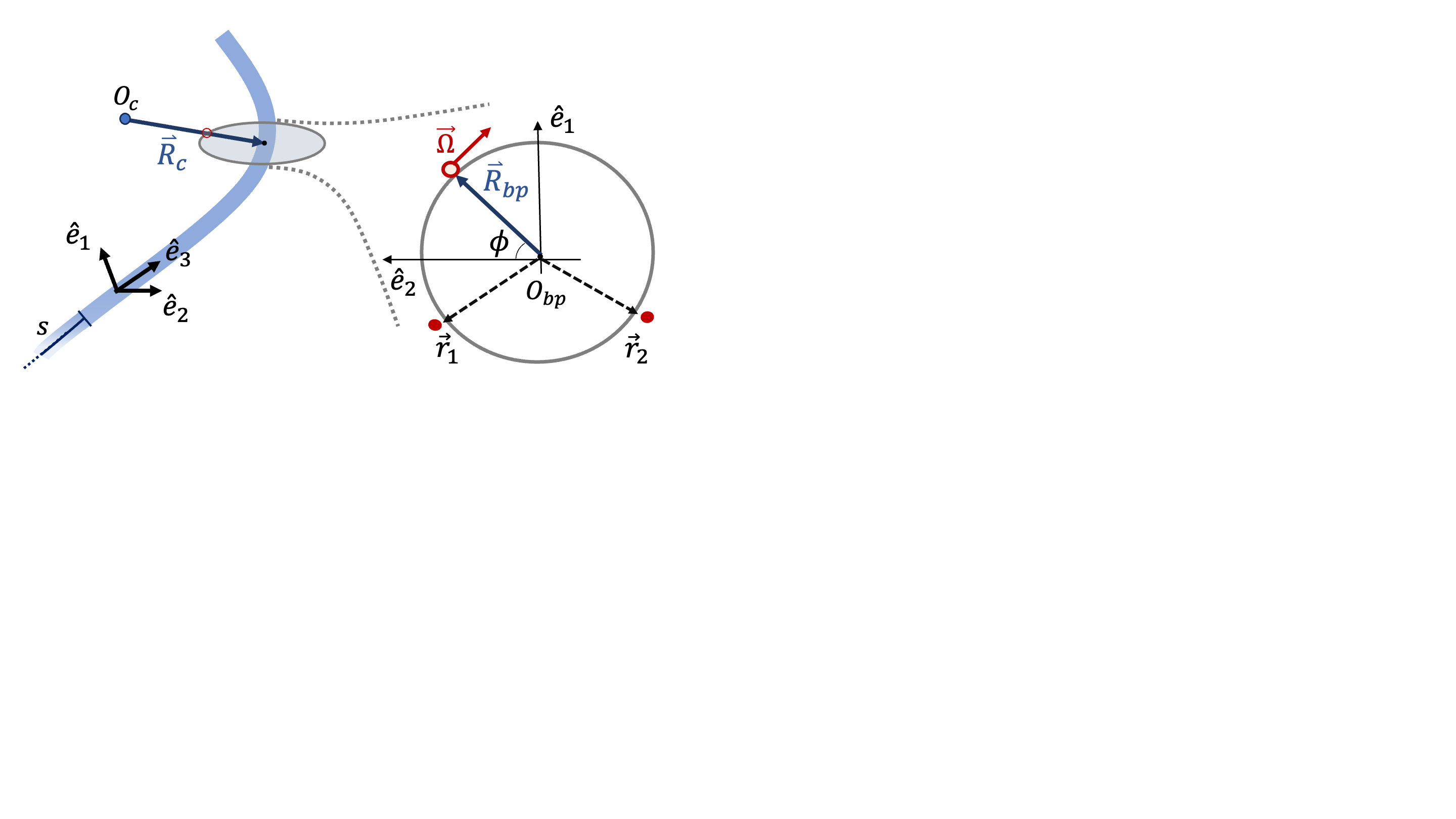}
\caption{Schematic figure of a cross section of a base pair in a curved strand. When $O_c$ and $\vec{R}_c$ are the center of the curvature and the radius of the curvature that the strand draws at the n th base pair, a hollow red circle indicates the contact point between $\vec{R}_c$ and the circumference of the cross section disk of a base pair. $O_{bp}$ in the right figure is the center of the cross section of the base pair, which has two red dots as the location of the nucleotides.  Reprinted figure with rearrangement by the authors and the permission from Nomidis et al.\cite{Nomidis2019b} and APS, doi:10.1103/PhysRevE.99.032414.}
\end{figure}

%some paragraph - why 3DNA? - how to measure?  
%There are many combination of a set of rotation vector $\vec{\Omega}(s_i)$ derived from various arrangement of sequence-dependent elasticity that draws the curvature of the strand $\Omega^s$.

When two nucleotides in a base pair lie at $\left(-r_{\rho},r_{\tau},0\right)$ and $\left(-r_{\rho},-r_{\tau},0\right)$ in the coordinate system $ \{\hat{\textbf e}_i\}_{n}$ as shown in Fig. 1, the vectors $\vec{r}_1$ and $\vec{r}_2$ from the center of cross section to each nucleotide can be defined respectively. The center of the cross section is defined along the arc length of the strand. Similarly, the stacking vector $\vec{\ell}$ is defined along $\hat{\textbf{e}}_3$ at $O_{bp}$ in Fig. 1. The deformation along roll ($\rho$), tilt($\tau$) and the stacking component is derived from the vector $\vec{r_{\rho}}$, $\vec{r_{\tau}}$ and $\vec{\ell}$, which are $\vec{r}_1+\vec{r}_2$, $\vec{r}_1-\vec{r}_2$ and the stacking distance $\ell$ along $ \hat{\textbf{e}}_3$ substituting $\vec{x}$ in Eq.(\ref{eq:eq2}), as followings: %$\vec{r}_1$ and $\vec{r}_2$:  

\begin{eqnarray}
%\frac{1}{2} \left( \frac{d\vec{r}_1}{ds} + \frac{d\vec{r}_2}{ds} \right). \begin{bmatrix}-r_\tau &  0 &  0 \end{bmatrix}^T
 {d\vec{\rho}}= \pmb{ \Theta} \vec{r_{\rho}}  = \begin{bmatrix} 0 & -(\Omega_3+\omega_0) r_{\rho} & \Omega_2  r_{\rho}\end{bmatrix}^Tds,\label{eq:eq3} \\
%\\  
%\frac{1}{2} \left( \frac{d\vec{r}_2}{ds} - \frac{d\vec{r}_1}{ds} \right) 
{d\vec{\tau}}=   \pmb{\Theta} \vec{r_{\tau}} =   \begin{bmatrix} (\Omega_3+\omega_0) r_{\tau}& 0  & -\Omega_1  r_{\tau}\end{bmatrix}^Tds, \label{eq:eq4}\\
%=.  \Theta \begin{bmatrix} 0 & r_\rho & 0 \end{bmatrix}
%\frac{d\vec{\ell}}{ds} = 
{d\vec{\ell}}=\pmb{\Theta} \vec{\ell}  =  \begin{bmatrix} \Omega_2 \ell & -\Omega_1 \ell & 0 \end{bmatrix}^Tds.  \label{eq:eq5}
%=  Theta \begin{bmatrix} 0 &  0 & \ell \end{bmatrix}
\end{eqnarray}
Each component along $\hat{e}_3$ in Eq.(\ref{eq:eq3}) and Eq.(\ref{eq:eq4}) becomes the deformation defined as roll($\rho$) and tilt($\tau$), respectively. $\hat{ \textbf e}_{1}$ component is equivalent to shift($D_x$) and $\hat{ \textbf e}_{2}$ component becomes slide($D_y$) in Eq.(\ref{eq:eq3})$\sim$Eq.(\ref{eq:eq5}). Note that there is no $\hat{\textbf{e}}_3$ component in Eq.(\ref{eq:eq5}). Therefore, no rise($D_z$) is expected. The median of the two nucleotides is perturbed by Eq.(\ref{eq:eq3}) and Eq.(\ref{eq:eq5}) due to the assumption of a rigid cross section for each base pair and the given helicity, $\omega_0$, of the curved strand.

 Any deformation between nucleotides is ignored. Consequently, the location of $O_{bp}$ is altered with

\begin{eqnarray}\label{eq:eq6}
 %\frac{1}{2} \left( \frac{d\vec{r}_1}{ds} + \frac{d\vec{r}_2}{ds} \right)+ \Delta \vec{L}
  d \vec{R}_{c} =   \begin{bmatrix} \Omega_2 \ell & -\Omega_1 \ell -(\Omega_3+\omega_0) r_{\rho} & 0 \end{bmatrix}^Tds.
\end{eqnarray}
The dislocation $\Delta \vec{R}_{c} = [\Delta R_x, \Delta R_y,0]^T$ of the center of the cross section of the base pair is quantified from the integration of Eq.(\ref{eq:eq6}) along the arclength. The altered location of the center of the cross section causes an additional twist ($\Delta \omega$) in the predefined helicity between the base pairs, $\omega_0$, as shown in Fig. 2A-a and A-b. This additional twist angle, $\Delta \theta_\omega$ is derived from the geometrical condition in the curved strand with $\Delta \vec{R}_c$. From the component of $\Delta \vec{R}_{c}$ as shown in Fig. 2A-c, the value of additional twist($\Delta \theta_\omega$) becomes

 \begin{eqnarray}\label{eq:eq8}
 \Delta \theta_\omega = tan^{-1} \frac{\Delta R_y}{-\Delta R_x + r_{\rho}}.
 \end{eqnarray} 
Each component of  $\Delta \vec{R}_c$ from Eq.(\ref{eq:eq6}) is derived from the roll($\rho$) and tilt($\tau$), which are derived from $\Omega_2$ and $\Omega_1$, respectively, according to the third component in Eq.($\ref{eq:eq3}$) and Eq.($\ref{eq:eq4}$).

\subsection{Quantification}

\begin{figure}
\includegraphics[scale=0.35, trim={0 100 450 20},clip]{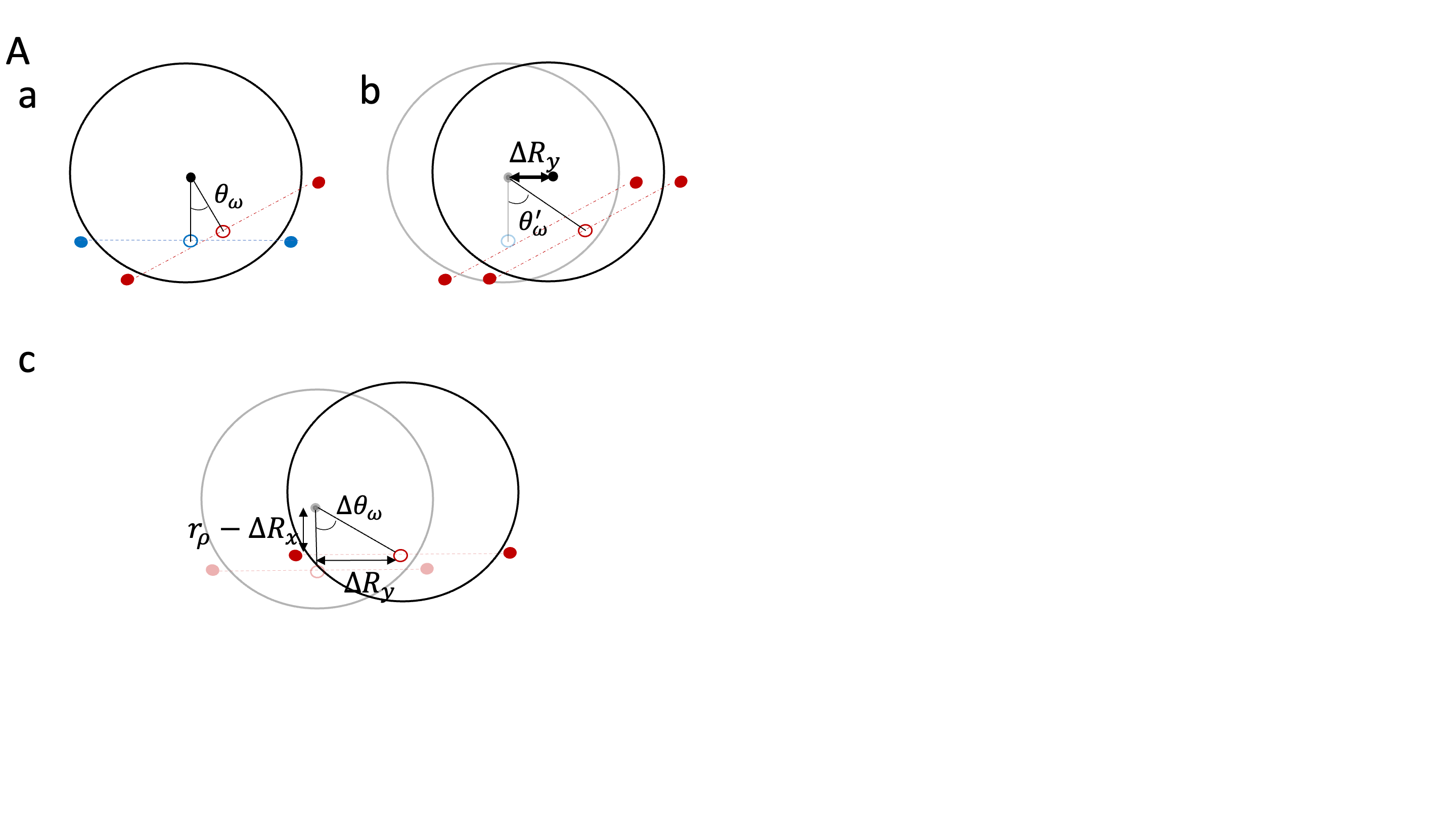}
\includegraphics[scale=0.35, trim={0 100 250 0},clip]{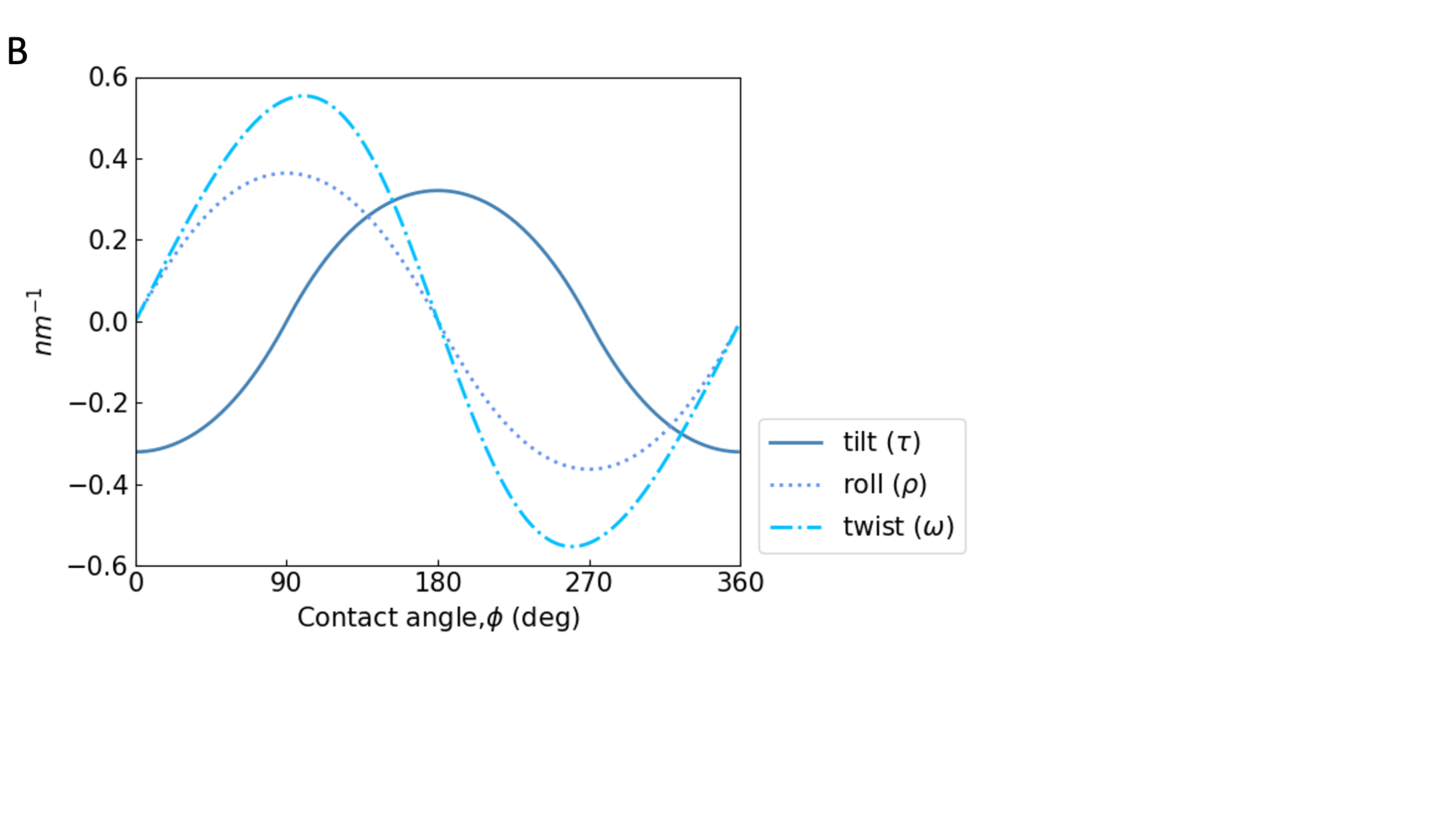}

\caption{ A-a. Arrangement between the $ n-1$th (blue) and the $ n$th (red) base pairs. $\theta_{\omega}$ is the angle for inherent helicity, which is from 32.4 $^{\circ}$. Two solid dots represent two nucleotides for each base pair, A-b. Modified twist angle, $\theta^{'}_{\omega}$ with $\Delta R_y$, A-c. Additional twist angle $\Delta \theta_{\omega}$ for $\theta^{'}_{\omega} = \theta_{\omega}+\Delta \theta_{\omega}$ with $\Delta R_y$ and $\Delta R_x$. Note that $\Delta \theta_{\omega}$ is measured from the median of the $ n$th base pair. B. Deformation of each rotational variable in curvature units along the contact angle. The maximum value of twist($\omega$) in the figure is around 10$^{\circ}$. }
\end{figure}

%Theoretical intricacies of bend-twist coupling in a helical system arise from the geometrically imposed condition for the ratio of the two bending components, which is governed by the cross-sectional orientation of the base pair and is dynamically altered by twist\cite{Marko1994}. If there is twist deformation induced by bend-twist coupling elasticity, or equivalently by the Frenet formula from differential geometry, it alters the ratio between the two bending components because the twist deformation affects the angle. An altered ratio of bending components introduces a new set of bend and twist from their coupling, and so on. 

The norm of the bending components is equivalent to the inverse of the radius of the curvature of the strand,  $(\Omega_1^2+\Omega_2^2)^{1/2}=1/|\vec{R}_c|$. The rotation axis of the bending components $(\Omega_1,\Omega_2)$ in $\vec{\Omega}$ lies along the tangent of the circumference of the cross section of the base pair on the contact point that defines the angle $\phi$ in Fig. 1.  It is, then, the angle $\phi$ in Fig. 1 that defines the ratio between bending components $(\Omega_1,\Omega_2)$. The result of each curvature deformation from Frenet-Serret formula\cite{DF3} and privous works\cite{Marko1994} becomes 

\begin{align}\label{eq:omega_phi}
\vec{\Omega}_{bp} = \left( \mp\Omega \cos\phi, \pm \Omega \sin\phi, \Omega_{\phi} \right). 
\end{align}
$\Omega_\phi$ is the twist deformation that is given from Eq.(\ref{eq:eq8}) with the bending components in Eq.(\ref{eq:omega_phi}). First and second term in Eq.(\ref{eq:omega_phi}) indicates the curvature formed by tilt($\tau$) and roll($\rho$), respectively. The double signs appear in the same order. From Eq. (\ref{eq:eq8}) expressed in terms of bending components, the preferred twist deformation is established as a geometric constraint for a rigid base-pair cross-section. Consequently, altering the twist($\omega$) following the result of Eq. (\ref{eq:eq8}) simultaneously generates a new set of bending components during the evolution of the Hamiltonian dynamics. Note that the sequence-dependent bend-twist coupling elasticity is not perfectly aligned with the geometric constraints in Eq. (\ref{eq:eq8}) and Eq.(\ref{eq:omega_phi}), thereby introducing significant theoretical complexity into the bend-twist coupling during the evolution process. This evolution derives the updated contact angle shown in Fig. 1, which alters the contact angle ($\phi $) in Eq. (\ref{eq:omega_phi}). %The order of counting Eq. (\ref{eq:eq8}) and Eq. (\ref{eq:omega_phi}) for this deduction is not critical, but it certainly complicates the strand dynamics and obscures clear quantification.  %The derivation of the bending component in the rotational vector, $\vec{\Omega}$, is not equivalent to the previous study\cite{Marko1994} for its alternation of $\pm$ sign, but the alternation of the sign can be a reasonable expression due to the fluctuation given from the sequence-dependentcoupling elasticity as mentioned in the previous section. 

 %When the additional twist defined in Eq.(\ref{eq:eq8}) is applied to $\omega_0$ in Eq.(\ref{eq:eq2}) as $\Omega_3$ in $\vec{\Omega}$, the contact angle $\phi$ is perturbed so as the component of $\pm \left( -\Omega \cos\phi, \Omega \sin\phi \right)$ which is equivalent to  $\Omega_1$ and $\Omega_2$ of $\vec{\Omega}$. However, 
 
 One condition that simplifies the complication is the fact that the perturbation of $\phi$ only changes the ratio of the bending components, not the norm of bending. Therefore, the bend-twist coupling under the geometrically given conditions in Eq. (\ref{eq:eq8}) can be quantified with a fixed strand curvature. In our case, the strand's curvature is fixed by the core structure's radius, while the ratio of bending components can vary. Then, the results of roll($\rho$), tilt($\tau$), and twist($\omega$) derived from the contact angle $\phi$ and the radius of curvature that wraps around an artificial spherical core structure with a radius of 3.5 $nm$, are shown in Fig. 2-B. As a result, the correlation derived from Eq.(\ref{eq:eq8}) and Eq.(\ref{eq:omega_phi}), as shown in Fig. 2-B, becomes an ansatz that is valid during the dynamical evolution process from Hamiltonian dynamics. The effect of the alteration in contact angle $\phi$ is discussed in the following paragraphs with a computational proof using coarse-grained molecular dynamics simulations.

%\subsection{Simulation Models}
%\subsubsection{Simulation condition}
%A superhelix of nucleosomal DNA is built with multiple protein attachments at different regions simultaneously \cite {Brandani2021}. 

%\section{Results}

\subsection{Bend-twist coupling in open-ended superhelix}

Two sets of strand information are adapted from Freeman et al.\cite{Freeman2014a}. A set of two strands, IAT and EXAT, shares four replacements in their sequences, marked by AGT in IAT and AAT in EXAT. In another set which has c1, c2, and c3 strands, a slightly different strategy is adapted. The number of TA sequence occurrences increases in the order of c1$<$c3$<$c2. They share the identical location of the replacement of TA sequence. All five strands are composed of 147 base pairs and calculated with the oxDNA1 and oxDNA2 packages\cite{Henrich2018,Snodin2015,Sulc2012,Ouldridge2011,Ouldridge2010}. The validation of Eq.(\ref{eq:eq3})$\sim$Eq.(\ref{eq:eq8}) is conducted using the oxDNA1 and oxDNA2 simulations to confirm the geometric constraints derived from the major-minor groove, which is well dictated in oxDNA2 only.  A charged spherical core structure, such as a nanoparticle (NP) with a diameter of 7 nm is used in the simulation. The simulation is conducted at 300 K with the potential energy and other details in the Supporting Information. 
 
  %The temperature profile and persistence length calculated with the proposed thermostat embedded in the oxDNA package are included in the Supporting Information, validating the robustness of our computational results. The wrapping conformation calculated with the Langevin thermostat is identical to that obtained with the heat diffusion thermostat. Yet, the wrapping formation does not extend to a time scale sufficient for analysis. 

% The quantification for the series of multiple attachments of the proteins is not considered in this paper. 

\begin{figure}
\includegraphics[scale=0.4, trim={0 120 0 0},clip]{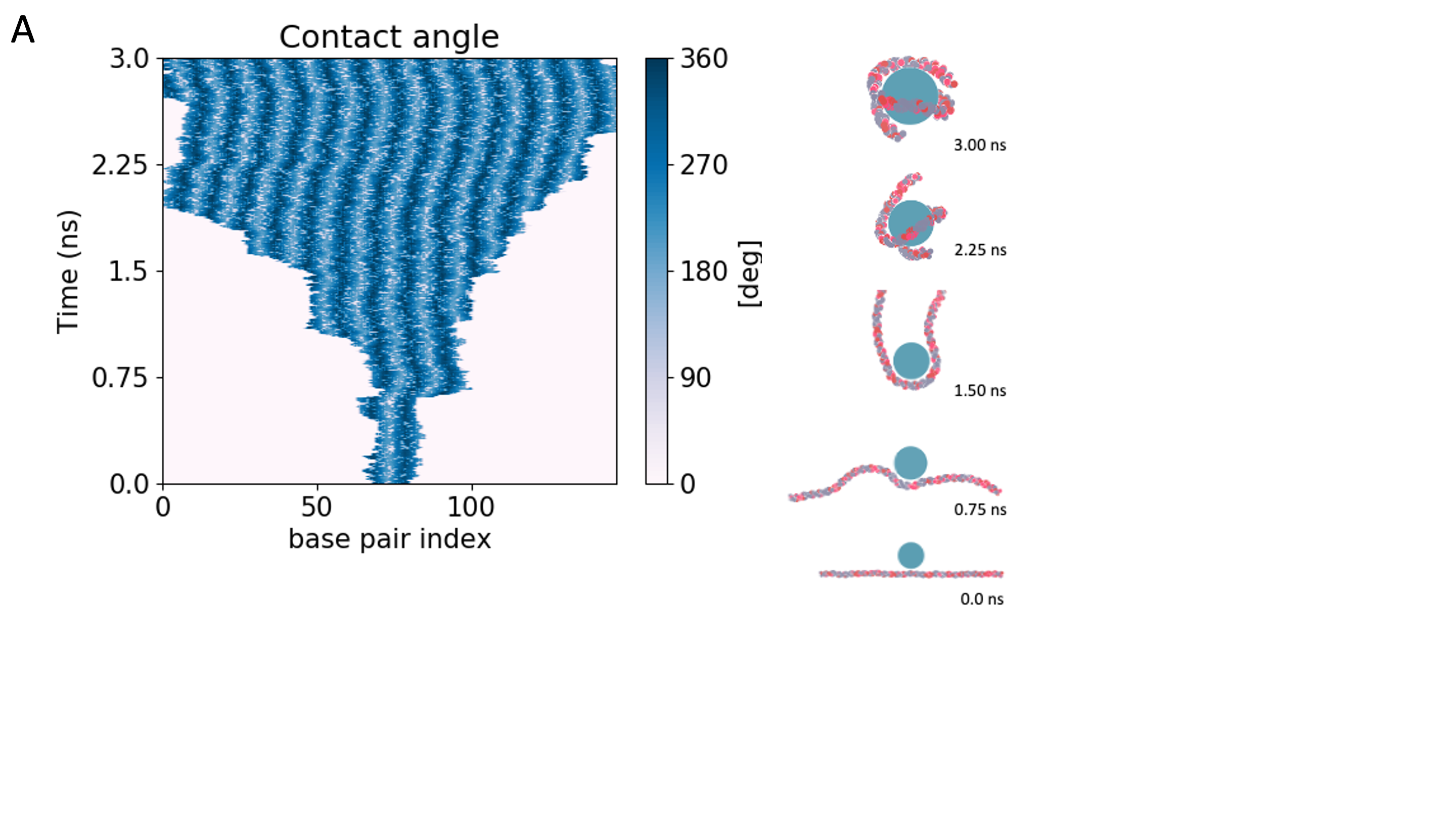}\\
\includegraphics[scale=0.4, trim={0 80 0 0},clip]{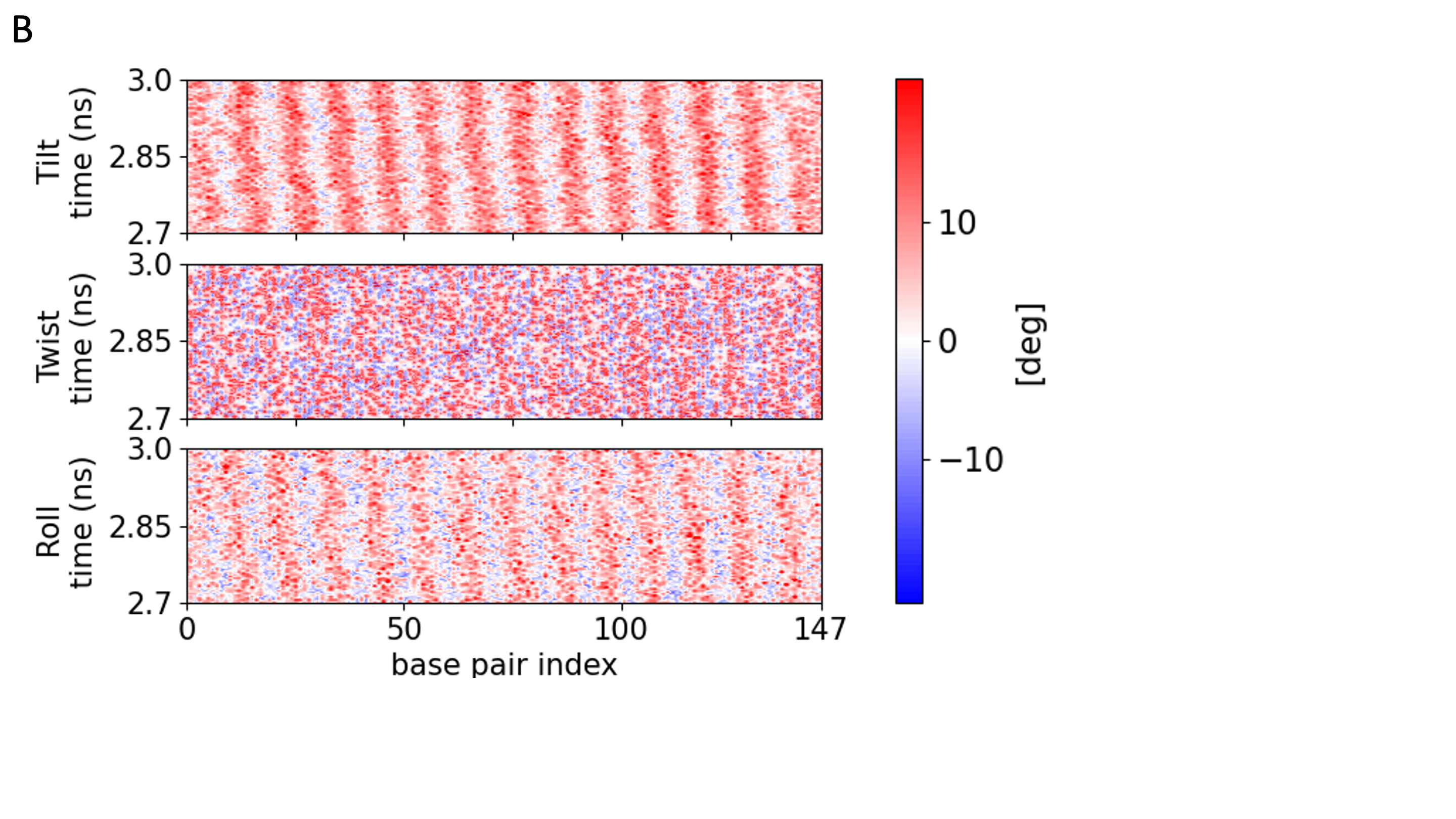}
\caption{A. The contact angle $\phi$ in spatiotemporal distribution. y axis is time, x axis is the index of base pair along the strands. The right inset is about the conformation change during the wrapping process of c1 strand. B. The spatiotemporal distributions for three rotational deformations along roll($\rho$), tilt($\tau$) and twist($\omega$) in 2.7 ns$\sim$3 ns. In the case of a twist, the blue and red contour lines are added for around -4 and 4 degrees, respectively, and the value of twist is the result of the total twist subtracted by 34 degrees to show a clear distribution. }
\end{figure}

The contact angle($\phi$) for each base pair of the c1 strand is calculated using the trajectory from oxDNA2 simulation with NP. The result shows the spatiotemporal distribution in Fig. 3. The clear pattern of $\phi$ shows the repetition from 0 to 360 degrees for each 10.3 turns in Fig. 3A along the strand during superhelix formation, as drawn in the right inset. The contact angle $\phi$ is measured only when the strand is within 6 nm of the center of the core structure. Subsequently, as shown in Fig. 3B, the curvature of each primary rotational variable shows a correlated pattern with the contact angle during the wrapping process from 2.7 ns to 3.0 ns. The spatiotemporal distributions of tilt($\tau$), twist($\omega$) and roll($\rho$) show the patterns as the derivation in Eq.(\ref{eq:eq4})$\sim$Eq.(\ref{eq:eq6}), which are functions of the contact angle($\phi$) with different phase each other. Note that the twist($\omega$) and roll($\rho$) share the same phase in the distribution, unlike tilt($\tau$). The animated GIF in Supporting Video SV1 provides the wrapping process of the c1 strand, which is shown in the inset of Fig. 3A. The remaining spatiotemporal distributions of the three rotational and three translational variables across all five strands during wrapping formation are included in the Supporting Information. 
%the ratio of roll($\rho$) and tilt ($\tau$) is also altered with additional twist, the angle $\phi$ should be ranged from 0$^{\circ}$ to 360$^{\circ}$ during 10.3 bps with a constant fluctuation . 

\subsection{Superhelix induced by bend-twist coupling}

 %The Darboux frame is defined along the tangent of the curvature so that Eq.(\ref{eq:eq6}) and Eq.(\ref{eq:eq8}) include the translational component of the deformation between base pairs from the Frenet-Serret formula, which depicts the location of the base pair in neighborhood. Among these translational deformations, 
 
 The relocation of the center of the base pair cross section in Eq.(\ref{eq:eq6}) drives a component perpendicular to the plane defined by $\hat{e}_3$ and the radius of curvature, $\vec{R}_c$, in Fig. 1. When there is rearrangement of the cross section perpendicular to the curvature plane, which is kurtosis, the curved strand produces a superhelix as in nucleosomal DNA. Superhelix formation from this geometrically imposed condition indicates that the curved strand automatically forms in a nucleosomal DNA-like conformation when sufficient curvature is provided. The base-pair-wise kurtosis of the open-ended strand can be derived from the sum of vectors defined by shift($D_x$) and slide($D_y$) using a coordinate transformation. With additional twist($\Delta \omega$) deformation in Eq.(\ref{eq:eq8}), each term of shift($D_x$) and slide($D_y$) becomes 

\begin{eqnarray}\label{eq:eq_tra}
\vec{D_x} = \ell^2 \Omega_2 \hat{\textbf e}_1,\nonumber\\
\vec{D_y}=r_{\tau} \ell ( \omega_0 +\Delta \omega) \hat{\textbf e}_2  - \ell^2 \Omega_1 \hat{\textbf e}_2. 
\end{eqnarray}

Eq.(\ref{eq:eq_tra}) can be quantified along the new unit axis $\hat{\textbf e}_{c}$, aligned along $\vec{R}_c$ in Fig. 1, and the axis $\hat{\textbf e}_{\mathcal K}$ for the kurtosis, which is normal to the plane defined by  $\hat{\textbf e}_{c}$ and $\hat{\textbf e}_3$. The coordinate transformation for kurtosis is defined with the angle $\phi$ as $\hat{ \textbf e}_{\mathcal{K}} = -sin \phi \hat{ \textbf e}_2 + cos \phi \hat{ \textbf e}_1$. The quantification of the kurtosis occurred between two base pairs, $\mathcal{K}_{bp}$ becomes
%unavoidable for the strand in a certain curvature with enough length for the translational deformation,
\begin{eqnarray} \label{eq:eq9}
\mathcal{K}_{bp} = D_x \cos \phi -  D_y \sin \phi,  \nonumber  \\
=  \pm\ell^2 \Omega_2  \cos \phi -(r_{\tau} \ell ( \omega_0 +\Delta \omega)  \mp \ell^2 \Omega_1) \sin \phi,  \nonumber  \\
= r_{\tau} \ell (\omega_0+\Delta \omega)\sin \phi \pm 2  \ell^2 \Omega \sin 2 \phi.   
\end{eqnarray}
The accumulation of $\mathcal{K}_{bp}$ in Eq.(\ref{eq:eq9}) along the strand is identical to the kurtosis that makes the strand form a superhelix. During 10.3 bp turns, the kurtosis involved with $\omega_0$ in the first term and the second term on the RHS of Eq.(\ref{eq:eq9}) becomes zero because $\phi$ varies from 0$^{\circ}$ to 360$^{\circ}$ in every 10.3 bp. In the meantime, $\Delta \omega \sin \phi$ in the first term on the RHS of Eq.(\ref{eq:eq9}) has a non-zero accumulation during 10.3 bp since $\Delta \omega$ shown in Fig. 2B has the same +/- sign as $\sin \phi$.

 From Eq.(\ref{eq:eq_tra}) and Eq.(\ref{eq:eq9}), we can confirm that the kurtosis remains with the additional twist deformation, $\Delta \omega$, that occurs from the bending, as derived by Eq.(\ref{eq:eq8}). The additional twist, $\Delta \omega$ as shown in Fig. 2B has its maximum value which is $\Delta \omega =$ 0.6 $nm^{-1}$, then the total kurtosis $\mathcal{K}$ becomes approximately 5.8 $nm$ with 147 bps, from the radius of the cross section of the base pair, 1 $nm$ and the stacking distance, $\ell =$ 3.4 $\text{\AA}$. The height of the nucleosomal DNA in the experiment is about 5.5 $nm$\cite{TJRichmond1984}.  The derivation is valid for any double helical structure with major and minor grooves where slide($D_y$) is non-zero. The range of real kurtosis could vary due to helicity alternation and any internal or external sources that affect the deformation or arrangement between nucleotides. From the derivation in Eq.(\ref{eq:eq_tra}) and Eq.(\ref{eq:eq9}), we confirm that the kurtosis $\mathcal{K}_{bp}$ is the key for the curved DNA strand to avoid self-contact in superhelix formation, and also the reason why the superhelical structure induced from supercoiling is geometrically stable.

\subsection{Role of Major-Minor Groove for 1.7 Turn }
\label{app6}

\begin{figure}
\includegraphics[scale=0.4, trim={0 300 0 0},clip]{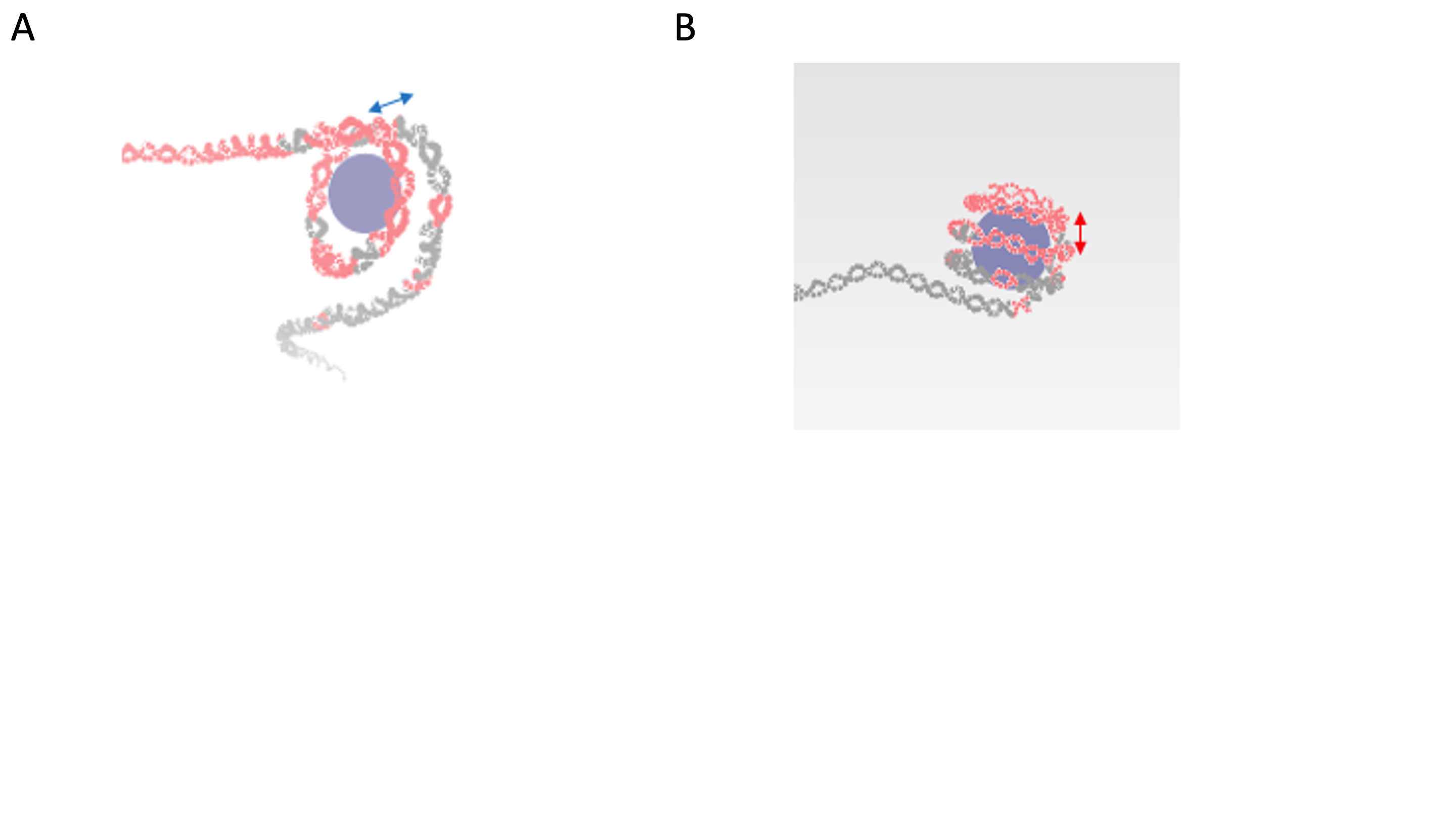}\\
\caption{A. The result of the wrapping process from the CG model with a major-minor groove(oxDNA2), B. from the CG model without a major-minor groove(oxDNA1). The distance between strands in the wrapping conformation(red arrow) in the oxDNA1 case is narrower than that calculated using oxDNA2(blue arrow). For the simulation, a 375 bp strand with an AT(red) and CG(gray) combination is used, with one end fixed. }
\end{figure}

For the additional twist, $\Delta \omega$ is dependent on the coordinates of nucleotides on the cross section as given by Eq.(\ref{eq:eq6}) and Eq.(\ref{eq:eq8}), the angle between nucleotides is the decisive feature for the kurtosis, $\mathcal{K}_{bp}$ in Eq.(\ref{eq:eq9}). When there is no difference between major and minor grooves, the angle between two nucleotides becomes 180 degrees. Two nucleotides are located at $\vec{r}_1=\left(-r_{bp},0,0\right)$ and $\vec{r}_2=\left(r_{bp},0,0\right)$ where $r_{bp}$ is the radius of the cross section of the base pair, $|\vec{R}_{bp}|$. Then, Eq. (\ref{eq:eq3}) becomes

\begin{eqnarray}\label{eq:eq11}
{d\vec{\rho}}= \pmb{ \Theta} \vec{r_{\rho}}  = 0. 
\end{eqnarray}

From the substitution of the angle $\theta$=120$^{\circ}$ between two nucleotides in Fig. 1 to $\theta=$180$^{\circ}$ makes the slide($D_y$)$\sim 0$ in Eq.(\ref{eq:eq_tra}). Therefore, the kurtosis $\mathcal{K}_{bp}$ and the integration of the kurtosis after the 10.3 bp turn become zero. 

This theoretically driven role of the major-minor groove is validated using oxDNA1 simulation, where coarse-grained particles compose two strands with $\theta=180 ^{\circ}$.  In oxDNA1, the condition of slide($D_y$) without major-minor groove asymmetry causes an increase in the number of wrappings compared to oxDNA2, as shown in Fig. 4A and B, since the self-contact is avoided due to the repulsion between coarse-grained particles. Conversely, oxDNA2 model with $\ell =$ 4.0$\text{\AA}$ calculates a slightly extended kurtosis of approximately 6.8 nm for a 147-bp strand. This configuration successfully completes 1.7 turns around a  NP with a diameter of 7 nm, as illustrated in Fig. 4A. The simulation results are presented in the animated GIF files included in Supporting Video SV2 (oxDNA1) and SV3 (oxDNA2). 

%\begin{gather}
%\begin{bmatrix}
%\hat{ \textbf e}_{\mathcal{R}} \\
%\hat{ \textbf e}_{\mathcal{K}}
%\end{bmatrix} = \Phi
%\begin{bmatrix}
%\hat{ \textbf e}_{\mathcal{2}} \\
%\hat{ \textbf e}_{\mathcal{1}}
%\end{bmatrix} = 
%\begin{bmatrix}
%cos \phi’ & -sin \phi’ \\
%sin \phi’ & cos \phi’ 
%\end{bmatrix}
%\begin{bmatrix}
%\hat{ \textbf e}_{2}\\
%\hat{ \textbf e}_{1}
%\end{bmatrix},
%\end{gather}

%with $\hat{ \textbf e}_{\mathcal{R}}$ is the unit vectors in the coordinate system defined along $\vec{R}_c$ and the normal vector of the cross section for the base pair, $\hat{e}_3$. The orthogonal vector defined between $\hat{ \textbf e}_{\mathcal{R}}$ and $\hat{ \textbf e}_3$ becomes the unit vector for the kurtosis $\hat{e}_{\mathcal{K}}$. With $\phi’=180-\phi$, we have $\hat{ \textbf e}_{\mathcal{K}} = -sin \phi \hat{ \textbf e}_2 + cos \phi \hat{ \textbf e}_1$.

\subsection{ Coupling elasticity in superhelix formation }
\label{app5}

Having fewer TA or AGT sequences in the c1 and EXAT strands is clearly beneficial for superhelix formation, as each simulation reaches the minimum free energy more rapidly than the other cases. The preference for superhelix formation, as marked with the free energy affinity in the legend of Fig. 5, is well reflected in the length of the duration of the wrapping process. Supporting Information includes further details on the method to quantify the affinity of the wrapping process.

%The elasticity measured using oxDNA2 with heat diffusion process, which is included in the Suppoting Information confirms that the energy contribution of the coupling elasticity between tilt($\tau$) - roll($\rho$) and tilt ($\tau$) - twist ($\omega$), which are $g_{\tau \rho}$ and $g_{\tau \omega}$, respectively, are insignificant compared to the coupling between twist($\omega$) and roll($\rho$), $g_{\omega \rho}$, which is consistent with the result from the free energy functional considering the symmetries\cite{Nomidis2019b,Skoruppa2017,Marko1994}. 
 The energy contribution of the coupling elasticity between tilt($\tau$) - roll($\rho$) and tilt ($\tau$) - twist ($\omega$), which are $g_{\tau \rho}$ and $g_{\tau \omega}$, respectively is insignificant compared to coupling between twist($\omega$) and roll($\rho$), $g_{\omega \rho}$. However, according to the geometrically given constraints,  $g_{\tau \rho}$ and $g_{\tau \omega}$ can regulate the speed and affinity of the wrapping process because the periodic structure of a double helix during 10.3 bp needs the full range of the contact points angle $\phi$, which is $ [0^{\circ},360^{\circ}]$ for the curvature formation. The set of curvature variables, which are roll($\rho$), tilt($\tau$), and twist($\omega$), is better to be in an identical pattern to the geometrically given constraints, as shown in Fig. 2B, for minimizing the deformation free energy.

 The combination of curvature variables in the free energy functional is preferably paired with the coupling elasticity corresponding to the phase designated in Fig. 2B, when wrapping occurs upon the ideal spherical core structure. While  $g_{\rho \omega }$ values remain largely unaffected by the replacements, $g_{\tau \rho }$ and $g_{\tau \omega }$ exhibit pronounced shifts in the number of instances satisfying the positive conditions $g_{\tau\rho} > 0$ or $g_{\tau\omega} > 0$. The total number of positive $g_{\rho\tau}$ in the replaced sequences is counted in Fig. 6, along with the affinity of the wrapping conformation for each strand in the set for c1/c2/c3. c1 has 20 $g_{\rho\tau}>0$ cases out of 24 base pairs that are affected by the replacement. c2 and c3 strands have less number of $g_{\rho\tau}>0$ compared to c1 strand. The high wrapping affinity and rapid wrapping speed of the c1 strand appear to underscore the importance of $g_{\rho\tau} > 0$, which indicates roll($\rho$) and twist($\omega$) are in phase as shown in Fig. 2B. All coupling elasticities, including neighboring parameters, are listed in Tables S4 to S11 of the Supporting Information. When AGT is replaced with AAT surrounded by cytosines, as shown in the IAT and EXAT cases, there are up to four times as many alternations in the elasticity followed by sequences, as shown in Table S3 in the Supporting Information. According to Fig. 2B, $g_{\tau\rho}$ and $g_{\tau \omega}$ are supposed to have positive and negative signs, respectively. This co-sign condition renders EXAT more suitable for forming curved strands compared to IAT. The lower number of $g_{\tau\rho} > 0$ cases also represents a key structural weakness in the IAT case, as shown in Table S3 in the Supporting Information. 
 
\begin{figure}
\includegraphics[scale=0.4, trim={0 200 0 0},clip]{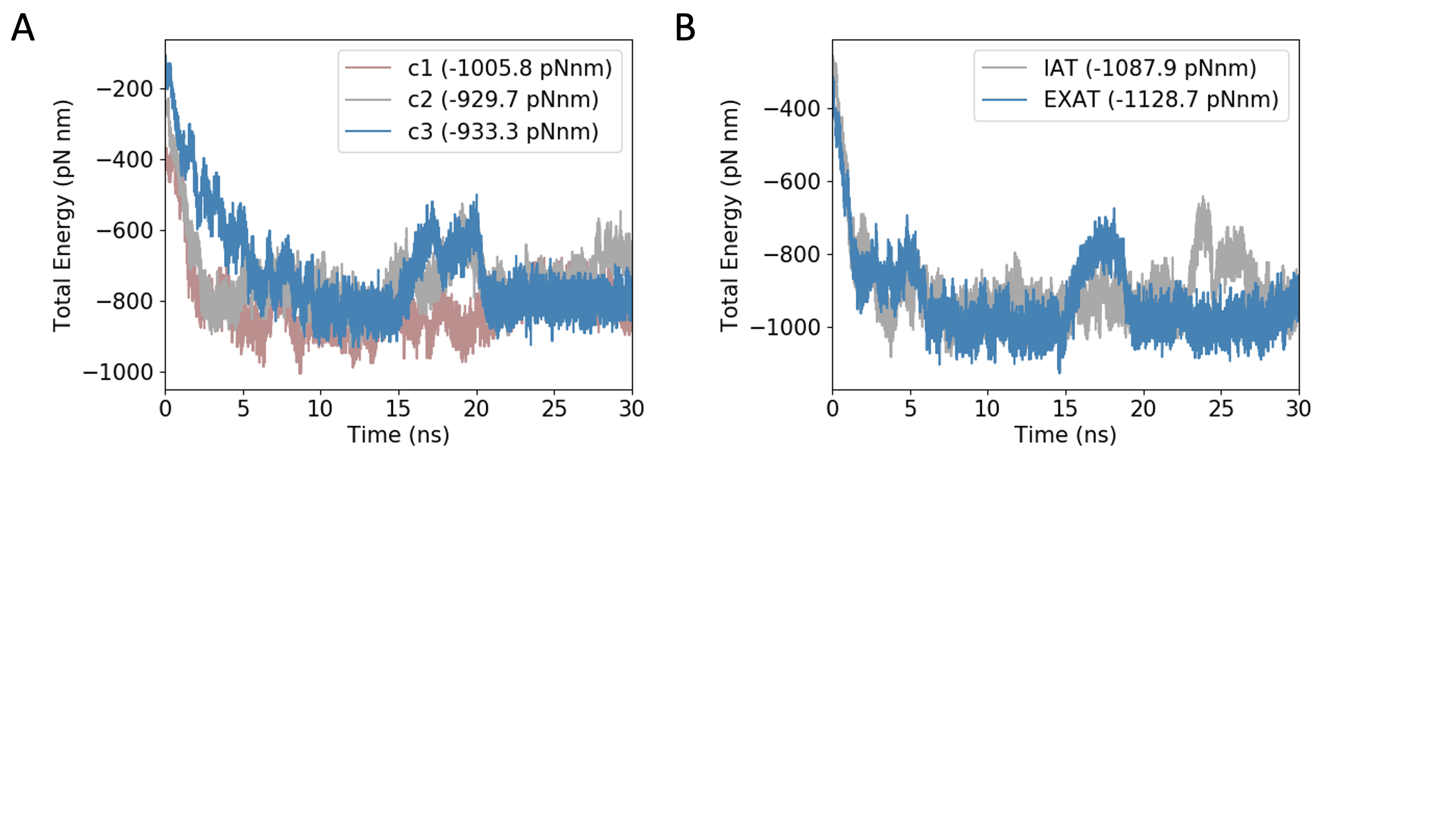}
\caption{Free energy trajectory during the wrapping process in 30 ns using oxDNA2 and a new thermostat. A. c1 (pink) proves its minimum free energy level and rapid completion of wrapping compared to c2(grey) and c3(blue) strands. B. EXAT(blue) has a more stable condition than IAT(grey) according to the minimum energy (in parentheses). }
\end{figure}

\begin{figure}
\centering
\includegraphics[scale=0.5]{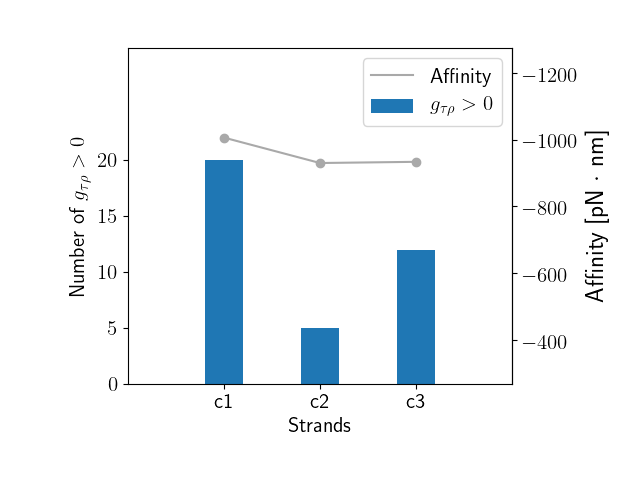}
\caption{Count of $g_{\tau\rho}>0$ sequences and affinity for c1, c2 and c3 strand.}
\end{figure}

In this study, the bare strand's bend-twist coupling, derived as mechanical and geometrical feature of the system, yields a different perspective on the stability of a curved B-DNA strand. The Frenet formula from differential geometry provides the shortest deformed bond length along the continuous trajectory that the molecule's conformation could take. Yet, a simplified derivation grounded in the Frenet formula of differential geometry provides key insights into the definition of the ansatz, elucidating the helical structure and its role in dynamical stability. With the geometrically given features in the dsDNA strand, the ansatz defines the base-pair-wise resolution bend-twist coupling derived from the Frenet-Serret formula by adjusting it with a set of vectors defined for the nucleotides in the base pair. The intercorrelation between the bending and twisting of the base pairs in the curved strand from the derivation is subsequently validated using oxDNA2 simulation.

 To counter the inherent fluctuations of helical strands or rods emerging from bend-twist coupling, the accumulation of deformation along the invariants of the ansatz provides a rigorous, base-pair-resolved metric for the stability of the curved DNA strand. The invariant from the ansatz is the non-zero distance drawn between strands in their supercoiling that makes the strand avoid self-contact. It is shown to be equivalent to the height of nucleosomal DNA, given the B-DNA geometry from the ansatz. This invariant is a geometrically given feature of the dsDNA strand, which is derived from major-minor grooves. Subsequently, the derivation clearly shows that there is no distance between strands to avoid self-contact when the angle between two nucleotides reaches 180 degrees, as confirmed by oxDNA1 and oxDNA2 simulations with different nucleotide arrangements\cite{Henrich2018,Sulc2012,Ouldridge2011,Ouldridge2010}. Finally, the stability of a strand formed under constant curvature originates from geometric features that align seamlessly with its underlying energetics, reinforcing the suggested mechanism of geometrically constrained stability within a chromatin fiber during an active translocation process in the presence of various sources of fluctuation. The analytical derivations and methodology presented in this work support a perspective based on helical strand geometry, which is directly applicable to the artificial design of helices, particularly for prescribing individual unit structure characteristics that govern dynamical stability.

%Given the range of bend-twist coupling observed in actin filaments within the range of thermal fluctuations \cite{enrique2010origin,borovikov2025twisting}, the artificially designed bend-twist coupling of synthetic helical systems could be tuned to a similar sensitivity. This sensitivity can offer a conditional activation cue for specific self-assembling structures or the regulation of assembly dynamics across various environments. Therefore, further approaches based on the dynamics characterized by geometrically driven ansatz offer intriguing possibilities for designing and modulating helical systems, whether artificial or natural, functioning either as active matter or regulators of emergent behavior.

\begin{acknowledgement}
The authors appreciate the fruitful discussion from Prof. Do-Nyun Kim to develop the manuscript.
This research is supported by Basic Science Research Program through the National Research Foundation of Korea(NRF) funded by the Ministry of Education (NRF-2020R1I1A1A01071567, NRF-2022R1I1A1A01063582) and  National Convergence Research of Scientific Challenges through the National Research Foundation of Korea(NRF) funded by Ministry of Science and ICT (NRF-2020M3F7A1094299). Its computational resources are from National Supercomputing Center with supercomputing resources including technical support (KSC-2020-CRE-0345).  There are no conflicts to declare. The authors thank Google Gemini for linguistic assistance and manuscript polishing.\end{acknowledgement}

%% Use figure environment to create figures
%% Refer following link for more details.
%% https://en.wikibooks.org/wiki/LaTeX/Floats,_Figures_and_Captions
%\begin{figure}[t]%% placement specifier
%% Use \includegraphics command to insert graphic files. Place graphics files in 
%% working directory.
%\centering%% For centre alignment of image.
%\includegraphics{example-image-a}
%% Use \caption command for figure caption and label.
%\caption{Figure Caption}\label{fig1}
%% https://en.wikibooks.org/wiki/LaTeX/Importing_Graphics#Importing_external_graphics
%\end{figure}


\begin{suppinfo}
 Supporting Information.pdf. The following files are available free of charge.

\begin{itemize}
  \item Filename: Supporting\_Information.pdf
  \item Filename: SV1\_c1\_wrapping.gif
  \item Filename: SV2\_ox1\_wrapping.gif 
  \item Filename: SV3\_ox2\_wrapping.gif 
\end{itemize}

\end{suppinfo}

\section{Data and Software Availability}
The data underlying this study are available in the published article, the Supporting Information. The code that procduce the data underlying this study is openly available in Github at \url{https://github.com/ieebon/DNA_dynamics}. 
%\end{Data and Software Availability}
\bibliography{Author_tex_.bib}

\end{document}